\documentclass[10pt,twocolumn,letterpaper]{article}

\usepackage{cvpr}              

\usepackage[accsupp]{axessibility}

\usepackage{graphicx}
\usepackage{amsmath}
\usepackage{amssymb}
\usepackage{amsbsy}
\usepackage{amsthm, amsfonts}
\usepackage{mathrsfs}
\usepackage{booktabs}

\usepackage{times}
\usepackage{bm}
\usepackage{setspace}
\usepackage{boldline,multirow}
\usepackage{makecell}
\usepackage[english]{babel}
\usepackage{xcolor}
\usepackage{pdfrender}

\usepackage[pagebackref,breaklinks,colorlinks]{hyperref}

\usepackage[capitalize]{cleveref}
\crefname{section}{Sec.}{Secs.}
\Crefname{section}{Section}{Sections}
\Crefname{table}{Table}{Tables}
\crefname{table}{Tab.}{Tabs.}

\def\cvprPaperID{4} 
\def\confName{CVPR}
\def\confYear{2023}

\definecolor{LimeGreen}{rgb}{0.2, 0.8, 0.2}

\begin{document}

\title{OpenFed: A Comprehensive and Versatile Open-Source Federated Learning Framework}

\author{Dengsheng Chen$^{1}$,  \hspace{2mm}
        Vince Junkai Tan$^{2}$,  \hspace{2mm}
        Zhilin Lu$^{3}$,  \hspace{2mm}
        Enhua Wu$^{4,5,6}$,  \hspace{2mm}
        Jie Hu$^{5,6}$\thanks{Corresponding author} \thanks{Project lead}
        \\
        
$^{1}$ Meituan  \hspace{5mm}
$^{2}$ Bytedance Inc.  \hspace{5mm}
$^{3}$ Tsinghua University  \hspace{5mm}
$^{4}$ University of Macau \\
$^{5}$ State Key Lab of Computer Science, ISCAS \hspace{5mm}
$^{6}$ University of Chinese Academy of Sciences \\
{\tt\small chendengsheng@meituan.com, vince.tan.jun.kai@gmail.com, luzl18@mails.tsinghua.edu.cn},\\
{\tt\small ehwu@um.edu.mo, hujie@ios.ac.cn}
}

\maketitle

\begin{abstract}
Recent developments in Artificial Intelligence techniques have enabled their successful application across a spectrum of commercial and industrial settings. However, these techniques require large volumes of data to be aggregated in a centralized manner, forestalling their applicability to scenarios wherein the data is sensitive or the cost of data transmission is prohibitive. Federated Learning alleviates these problems by decentralizing model training, thereby removing the need for data transfer and aggregation. To advance the adoption of Federated Learning, more research and development needs to be conducted to address some important open questions. In this work, we propose OpenFed, an open-source software framework for end-to-end Federated Learning. OpenFed reduces the barrier to entry for both researchers and downstream users of Federated Learning by the targeted removal of existing pain points. For researchers, OpenFed provides a framework wherein new methods can be easily implemented and fairly evaluated against an extensive suite of benchmarks. For downstream users, OpenFed allows Federated Learning to be plugged and play within different subject-matter contexts, removing the need for deep expertise in Federated Learning. The source code of OpenFed is publicly available online at \url{https://github.com/FederalLab/OpenFed}.
\end{abstract}

\section{Introduction}

Recent developments in Artificial Intelligence~(AI) and Machine Learning~(ML) techniques have advanced the state-of-the-art across many application domains~\cite{pouyanfar2018survey}.
In particular, Deep Learning~(DL) has revolutionized the field and deep neural networks are now the defacto standard for computer vision~\cite{feng2019computer}, natural language processing~\cite{otter2020survey}, audio and speech processing~\cite{zhu2021deep}, and reinforcement learning~\cite{arulkumaran2017deep}, among others. 
In the broader historical context of AI research, deep learning simply represents the next milestone in the Bitter Lesson~\footnote{http://www.incompleteideas.net/IncIdeas/BitterLesson.html} - ``general methods that leverage computation are ultimately the most effective''. 
Deep learning enables today's vast computational resources to be unleashed on equally vast amounts of data, with the best-known example perhaps being ImageNet's 14 million hand-annotated images~\cite{russakovsky2015imagenet}.
The typical setup involves centralizing the said data onto a database, which is connected to a high-performance server or cluster of servers. 
The training process can then take place, and the deep neural network is exposed to these examples and their corresponding desired outputs and therefore ``learns'' the assigned task.

While this paradigm has proven to be broadly effective, there are some use cases where it is not straightforwardly applicable~\cite{lim2020federated,aledhari2020federated}. 
The very first step of centralizing the data onto a database can be challenging or impossible for two main reasons: the data is personal or confidential and therefore the user would not consent to it being transmitted, or the data originates from edge devices and transmission costs are prohibitive.

Federated Learning (FL) is a family of ML techniques proposed to address these challenges~\cite{mcmahan2017communication}. 
Rather than centralizing the data and then training the model on a server, in FL, model training is decentralized to the data sources themselves, obviating data transfer entirely~\cite{wang2020federated,yu2020fairness}. 
Although FL has already been applied successfully in some industrial and commercial settings, it is in fact still an area of active research. 
Many open sub-problems need to be addressed before FL can be considered ready for broader application~\cite{li2020federated}. 

As a research topic, FL suffers from several stumbling blocks. 
Firstly, fair comparisons are difficult to perform.
Much research is built upon synthetically generated datasets. 
For these datasets, typically only the statistical properties can be practically reported. 
This introduces a significant element of randomness to experimental results. 
Secondly, in applied ML research, it is common to make use of existing frameworks which may not have been built with FL in mind. 
Because these frameworks are industry standards for their respective domains, it is not feasible to abandon them just to make use of FL techniques. 
Finally, to be effective in FL research requires a broad skill set. Whether to replicate existing work or to propose new methods, researchers need strong software engineering skills to implement the (pseudo) distributed setting that FL takes place in, a deep understanding of the ML principles that undergird FL techniques, as well as subject-matter expertise in the domain in question.

In this work, we propose OpenFed, a software framework that targets the stumbling blocks identified above in order to accelerate FL research and development. 
For FL researchers, OpenFed provides comprehensive implementations of existing methods and their corresponding performances across a suite of standardized datasets, thereby enabling convenient and fair comparisons for newly proposed methods while reducing the effort required to implement them. 
For FL users, OpenFed's support for common third-party frameworks allows state-of-the-art FL techniques to be integrated into practical applications without the need for practitioners to fully understand the underlying implementation. 

\section{Aims and contributions}

For both academic and industrial applications of FL, it is essential that the software framework used is versatile and flexible. 
There already exist several influential FL frameworks, each with its own focus. 
However, these frameworks have left some challenges unmet.

Firstly, the full diversity of FL topologies is not commonly supported. 
For example, TensorFlow Federated~\cite{ingerman2019introducing}, PySyft~\cite{ziller2021pysyft}, and LEAF~\cite{caldas2018leaf} support only the centralized topology.
Furthermore, these libraries do not provide convenient interfaces for the flexible exchange of auxiliary information or the customization of the training procedure, thereby limiting the algorithmic innovations possible.

Secondly, it is challenging to perform fair comparisons. 
Newly proposed FL algorithms are often implemented based on different libraries and dataset configurations. 
As a result, it is difficult for researchers to compare the performances of the algorithms fairly. 

Finally,  it is not easy to transition FL algorithms to applied scenarios. Existing FL frameworks have not sufficiently catered for downstream integration. 
For instance, it is not trivial to employ even the famous FedML~\cite{he2020fedml} to train models from well-established libraries like HuggingFace~\cite{wolf-etal-2020-transformers}.

\begin{table*}[!htp]
\renewcommand\arraystretch{2}
\resizebox{\linewidth}{!}{
\begin{tabular}{cl|cccccccccc}
&  & \makecell[c]{\Large CrypTen \\[5pt] \Large \cite{gunning2019crypten}\\[5pt]} & \makecell[c]{\Large FATE \\[5pt] \Large \cite{yang2019federated}\\[5pt]} & \makecell[c]{\Large PaddleFL \\[5pt] \Large \cite{ma2019paddlepaddle}\\[5pt]} & \makecell[c]{\Large PySyft \\[5pt] \Large \cite{ziller2021pysyft}\\[5pt]} & \makecell[c]{\Large FedML \\[5pt] \Large \cite{he2020fedml}\\[5pt]} & \makecell[c]{\Large FFL-ERL \\[5pt] \Large \cite{ulm2018functional}\\[5pt]} & \makecell[c]{\Large LEAF \\[5pt] \Large \cite{caldas2018leaf}\\[5pt]} & \makecell[c]{\Large TensorIO \\[5pt] \Large \cite{tensorio} \\[5pt]} & \makecell[c]{\Large TFF \\[5pt] \Large \cite{ingerman2019introducing}\\[5pt]} & \makecell[c]{\Large OpenFed \\[5pt] \Large (ours) \\[5pt]} \\ \clineB{1-12}{4}
\multicolumn{1}{c|}{\multirow{3}{*}{\begin{tabular}[c]{@{}c@{}} \\[-30pt] \LARGE Diverse \\[-5pt] \LARGE computing\\[-5pt] \LARGE paradigms\end{tabular}}}    & \Large Standalone simulation                                                              &         & \textcolor{LimeGreen}{\Large $\checkmark$} & \textcolor{LimeGreen}{\Large $\checkmark$}  & \textcolor{LimeGreen}{\Large $\checkmark$} & \textcolor{LimeGreen}{\Large $\checkmark$} &         & \textcolor{LimeGreen}{\Large $\checkmark$} &          & \textcolor{LimeGreen}{\Large $\checkmark$} & \textcolor{LimeGreen}{\Large $\checkmark$} \\ \cline{2-12} 
\multicolumn{1}{c|}{}                                                                                            & \Large Distributed computing                                                             & \textcolor{LimeGreen}{\Large $\checkmark$} & \textcolor{LimeGreen}{\Large $\checkmark$} & \textcolor{LimeGreen}{\Large $\checkmark$}  & \textcolor{LimeGreen}{\Large $\checkmark$} & \textcolor{LimeGreen}{\Large $\checkmark$} &         &         &          & \textcolor{LimeGreen}{\Large $\checkmark$} & \textcolor{LimeGreen}{\Large $\checkmark$} \\ \cline{2-12} 
\multicolumn{1}{c|}{}                                                                                            & \Large On-device training                                                                 &         &         &          &         & \textcolor{LimeGreen}{\Large $\checkmark$} & \textcolor{LimeGreen}{\Large $\checkmark$} &         & \textcolor{LimeGreen}{\Large $\checkmark$}  &         & \textcolor{LimeGreen}{\Large $\checkmark$} \\ \clineB{1-12}{4}
\multicolumn{1}{c|}{\multirow{4}{*}{\begin{tabular}[c]{@{}c@{}} \LARGE Flexible \\ \LARGE \& generic \\ \LARGE API design\end{tabular}}} & \Large Message flow                                                                       &         &         &          &         & \textcolor{LimeGreen}{\Large $\checkmark$} &         &         &          &         & \textcolor{LimeGreen}{\Large $\checkmark$} \\ \cline{2-12} 
\multicolumn{1}{c|}{}                                                                                            & \begin{tabular}[c]{@{}l@{}}\Large Customize data distribution\end{tabular} &         &         &          &         &         &         & \textcolor{LimeGreen}{\Large $\checkmark$} &          &         & \textcolor{LimeGreen}{\Large $\checkmark$} \\ \cline{2-12} 
\multicolumn{1}{c|}{}                                                                                            & \Large Unified paradigm                                                                  &         & \textcolor{LimeGreen}{\Large $\checkmark$} & \textcolor{LimeGreen}{\Large $\checkmark$}  & \textcolor{LimeGreen}{\Large $\checkmark$} &         &         &         &          & \textcolor{LimeGreen}{\Large $\checkmark$} & \textcolor{LimeGreen}{\Large $\checkmark$} \\ \cline{2-12} 
\multicolumn{1}{c|}{}                                                                                            & \Large Third-party support                                                               &         &         &          & \textcolor{LimeGreen}{\Large $\checkmark$} &         &         &         &          &         & \textcolor{LimeGreen}{\Large $\checkmark$} \\  \clineB{1-12}{4}
\multicolumn{1}{c|}{\multirow{5}{*}{\begin{tabular}[c]{@{}c@{}} \LARGE Topology\\ \LARGE customization\end{tabular}}}           & \Large Split                                                                              &         &         & \textcolor{LimeGreen}{\Large $\checkmark$}  & \textcolor{LimeGreen}{\Large $\checkmark$} & \textcolor{LimeGreen}{\Large $\checkmark$} &         &         &          &         & \textcolor{LimeGreen}{\Large $\checkmark$} \\ \cline{2-12} 
\multicolumn{1}{c|}{}                                                                                            & \Large Vertical                                                                           &         & \textcolor{LimeGreen}{\Large $\checkmark$} & \textcolor{LimeGreen}{\Large $\checkmark$}  &         & \textcolor{LimeGreen}{\Large $\checkmark$} &         &         &          &         & \textcolor{LimeGreen}{\Large $\checkmark$} \\ \cline{2-12} 
\multicolumn{1}{c|}{}                                                                                            & \Large Hierarchical                                                                       &         &         &          &         &         &         &         &          &         & \textcolor{LimeGreen}{\Large $\checkmark$} \\ \cline{2-12} 
\multicolumn{1}{c|}{}                                                                                            & \Large Decentralized                                                                      &         &         &          &         & \textcolor{LimeGreen}{\Large $\checkmark$} &         &         &          &         & \textcolor{LimeGreen}{\Large $\checkmark$} \\ \cline{2-12} 
\multicolumn{1}{c|}{}                                                                                            & \Large Centralized                                                                        & \textcolor{LimeGreen}{\Large $\checkmark$} & \textcolor{LimeGreen}{\Large $\checkmark$} & \textcolor{LimeGreen}{\Large $\checkmark$}  & \textcolor{LimeGreen}{\Large $\checkmark$} & \textcolor{LimeGreen}{\Large $\checkmark$} & \textcolor{LimeGreen}{\Large $\checkmark$} & \textcolor{LimeGreen}{\Large $\checkmark$} & \textcolor{LimeGreen}{\Large $\checkmark$}  & \textcolor{LimeGreen}{\Large $\checkmark$} & \textcolor{LimeGreen}{\Large $\checkmark$} \\  \clineB{1-12}{4}
\multicolumn{1}{c|}{\multirow{4}{*}{\begin{tabular}[c]{@{}c@{}} \LARGE Federated \\ \LARGE optimizer\end{tabular}}}             & \Large Aggregation customization                                                          &         &         &          &         & \textcolor{LimeGreen}{\Large $\checkmark$} &         &         &          &         & \textcolor{LimeGreen}{\Large $\checkmark$} \\ \cline{2-12} 
\multicolumn{1}{c|}{}                                                                                            & \Large Gradient accumulation                                                               &         &         &          &         &         &         &         &          &         & \textcolor{LimeGreen}{\Large $\checkmark$} \\ \cline{2-12} 
\multicolumn{1}{c|}{}                                                                                            & \Large Parameter penalization                                                             &         &         &          &         & \textcolor{LimeGreen}{\Large $\checkmark$} &         &         &          &         & \textcolor{LimeGreen}{\Large $\checkmark$} \\ \cline{2-12} 
\multicolumn{1}{c|}{}                                                                                            & \Large State synchronization                                                             &         &         &          &         &         &         &         &          &         & \textcolor{LimeGreen}{\Large $\checkmark$} \\  \clineB{1-12}{4}
\multicolumn{1}{c|}{\multirow{4}{*}{\LARGE Benchmarks}}                                                                 & \Large Computer vision                                                                    &         & \textcolor{LimeGreen}{\Large $\checkmark$} & \textcolor{LimeGreen}{\Large $\checkmark$}  & \textcolor{LimeGreen}{\Large $\checkmark$} & \textcolor{LimeGreen}{\Large $\checkmark$} &         & \textcolor{LimeGreen}{\Large $\checkmark$} & \textcolor{LimeGreen}{\Large $\checkmark$}  & \textcolor{LimeGreen}{\Large $\checkmark$} & \textcolor{LimeGreen}{\Large $\checkmark$} \\ \cline{2-12} 
\multicolumn{1}{c|}{}                                                                                            & \Large Natural language processing                                                        &         & \textcolor{LimeGreen}{\Large $\checkmark$} & \textcolor{LimeGreen}{\Large $\checkmark$}  &         & \textcolor{LimeGreen}{\Large $\checkmark$} &         & \textcolor{LimeGreen}{\Large $\checkmark$} &          & \textcolor{LimeGreen}{\Large $\checkmark$} & \textcolor{LimeGreen}{\Large $\checkmark$} \\ \cline{2-12} 
\multicolumn{1}{c|}{}                                                                                            & \Large Reinforcement learning                                                             &         &         &          &         &         &         &         &          &         & \textcolor{LimeGreen}{\Large $\checkmark$} \\ \cline{2-12} 
\multicolumn{1}{c|}{}                                                                                            & \Large Medical analysis                                                                   &         &         &          & \textcolor{LimeGreen}{\Large $\checkmark$} &         &         &         &          &         & \textcolor{LimeGreen}{\Large $\checkmark$} \\  \clineB{1-12}{4}
\end{tabular}
}
\caption{Comparison between OpenFed and existing FL frameworks.}
\label{tab:platform-summary}
\end{table*}

In order to address the above challenges, we designed a novel FL framework named OpenFed based on the PyTorch~\cite{paszke2019pytorch} ecosystem. 
OpenFed provides an expandable toolkit for FL algorithm development across diverse topologies. 
Comparisons between OpenFed and existing mainstream FL libraries are given in Table \ref{tab:platform-summary}. 
The main contributions of the OpenFed framework are listed as follows.

\paragraph{Diverse topologies} 
OpenFed provides powerful automatic topology analysis and construction tools. 
By introducing the concept of a federated group, we are able to decompose an entire FL topology into its atomic units. 
Novel FL algorithms for split, vertical, hierarchical, or decentralized topologies can then be implemented as though they were the standard centralized topology. 
In addition, OpenFed provides an interface for auxiliary information to be exchanged across compute nodes.

\paragraph{Comprehensive and standardized FL algorithms and benchmarks} 
OpenFed provides standardized datasets, algorithms, and benchmarks. 
Researchers proposing novel algorithm designs can conveniently and comprehensively compare their new ideas against existing solutions. 
OpenFed also provides rich configuration possibilities (e.g. server/client optimizer, sampling strategy of partial-activated clients, non-i.i.d. distribution of dataset partition), allowing researchers to better validate how well different algorithms generalize to specific situations.

\paragraph{User-friendly API design} 
Mainstream deep learning (DL) frameworks, such as PyTorch and TensorFlow, lack ready-made APIs for FL. 
OpenFed carefully inherits the API design from the PyTorch library, which is commonly used by the DL community. 
This reduces the barrier to entry for the larger DL community to participate in FL research and development. 
Furthermore, OpenFed provides rich built-in features including weighted gradient descent, federated averaging, diverse data augmentation, local early stopping, etc. 
These FL modules offer an immersive development experience to OpenFed users.

\paragraph{Third-party library support} 
The relative maturity of DL has resulted in the rise and success of open-source secondary libraries. 
These libraries provide state-of-the-art implementations for their respective domain areas. 
For example, HuggingFace~\cite{wolf-etal-2020-transformers} for natural language processing, MONAI~\cite{monai} for medical analysis, and MMCV~\cite{mmcv} for computer vision. 
However, few of this support FL natively. 
Furthermore, because these libraries are increasingly widely adopted, it is unlikely for newly proposed libraries that offer only FL as an advantage to gain much traction. 
Therefore, OpenFed is designed to be plug-and-play compatible with these third-party libraries.

\section{Architecture design}

\begin{figure*}[t]
  \centering
  \includegraphics[width=0.95\textwidth]{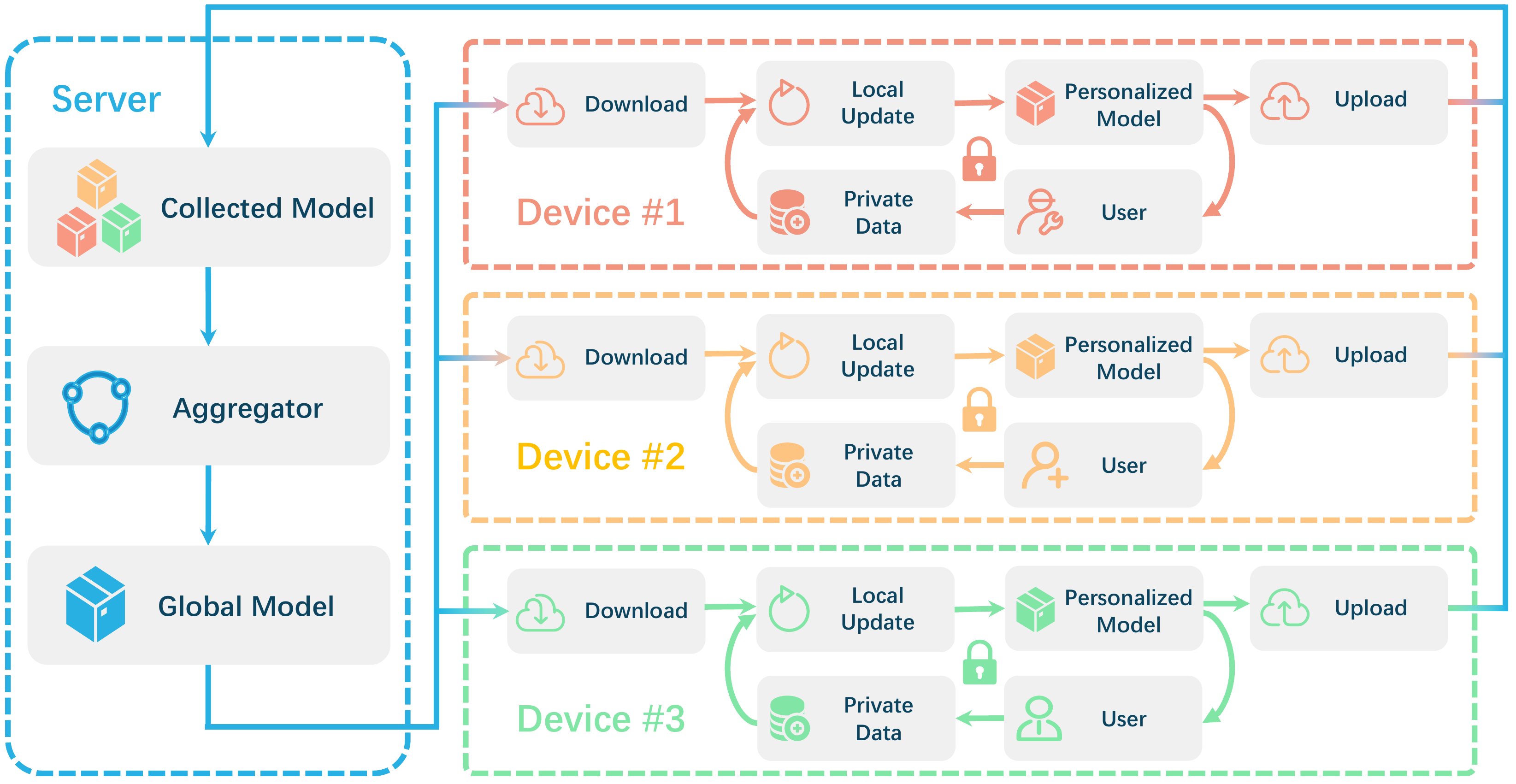}
  \caption{An illustration of the FL workflow based on the OpenFed framework. A server collects personalized models from different users, and the aggregator combines these to create a global model. This model is sent to all the users and is separately personalized on each device. The cycle then repeats, making up the main body of the FL workflow. Note that private data is only ever accessed by the local device to guarantee user privacy.}
  \label{fig:phase}
\end{figure*}

As depicted in Figure \ref{fig:phase}, the FL workflow is based on the distributed computing setup~\cite{yedder2021deep}, with a server and a number of local devices. 
Since user data never leaves the local device, the FL workflow inherently protects personal privacy.

Each local device downloads a copy of the most recent global model. 
This model can then be independently personalized using private data, resulting in improved performance for the user. 
These personalized models are then uploaded from local devices to the main server.
An aggregator at the server fuses all the personalized models into a newer and better-performing global model. 
Similarly, the on-device personalized model is also continuously improving with every round of this FL training loop. 
Note that the prediction task is typically also independently conducted on each local device.

\begin{figure*}[t]
  \centering
  \includegraphics[width=0.8\textwidth]{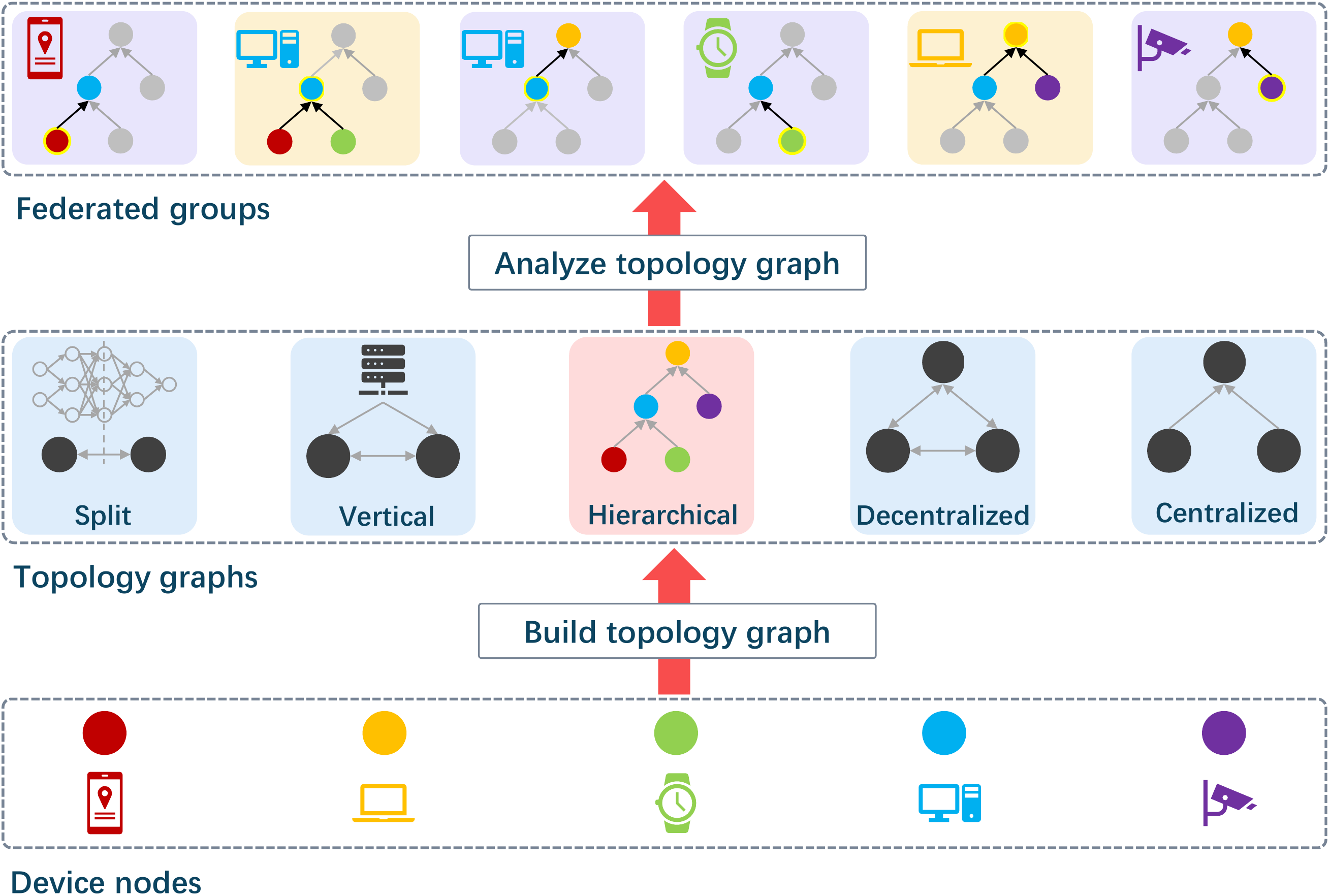}
  \caption{Three-level abstraction of the OpenFed design. Bottom level: nodes abstracted from various physical devices. Middle level: topology graph generated from the connection between different nodes. Top-level: federated groups deriving from the specific topology. Although one device might belong to several federated groups, communication among different devices is limited to within each group.}
  \label{fig:arch}
\end{figure*}

We designed the OpenFed architecture based on the FL workflow described earlier. 
The crucial architectural features of OpenFed are as follows.

\paragraph{Flexible federated groups abstraction} 
The connections between servers and local devices in an FL system can be complicated. 
Furthermore, the effort to implement these connections is often orthogonal to the algorithmic work that researchers primarily deal with.
Therefore, OpenFed includes an automatic topology analysis strategy in which different FL connections are expressed by the same atomic unit which is called a federated group. 
As Figure \ref{fig:arch} shows, the proposed strategy has three levels. 
Device nodes are added to the topology graph based on their connection modes. 
The topology graph is then parsed into a set of federated groups. 
The key insight is that every topology - even ones that appear highly irregular or complex - can be decomposed into a set of simpler groups that are individually just centralized topologies. 
By this mechanism, OpenFed enables centralized FL algorithms to be applied to other topologies.

\paragraph{The robust distributed training scheme} 
FL algorithms naturally assume a distributed training scenario since the servers and the devices are physically separated. 
OpenFed offers an interface to simulate dozens of federated groups on a single machine. 
This lowers the resource requirement to participate in and contribute to FL research. 
Furthermore, the distributed design in OpenFed allows complicated auxiliary information exchange within each federated group. 
For example, algorithmic hyper-parameters like learning rate, the FL round number, training instance, etc. can be easily shared. 
This enables the implementation of even highly exotic algorithms.

\paragraph{Specialized functional modules} 
The OpenFed architecture is composed of several functional modules. 
Researchers can pick and choose corresponding modules as required for their experiments. 
By using OpenFed's pre-defined modules like the aggregator, collaborator, pipe and optimizer, FL research can be accelerated. 
In addition, security and privacy-related functional modules are also supported in OpenFed, implementing defenses against common attacks like data and model poisoning \cite{nuding2020poisoning,feng2020practical}.

\section{Real-life application scenarios and benchmarks}

\begin{figure*}[htb]
\centering
    \hspace{20pt}
    \begin{subfigure}[b]{0.43\textwidth}
    \centering
        \includegraphics[width=\textwidth]{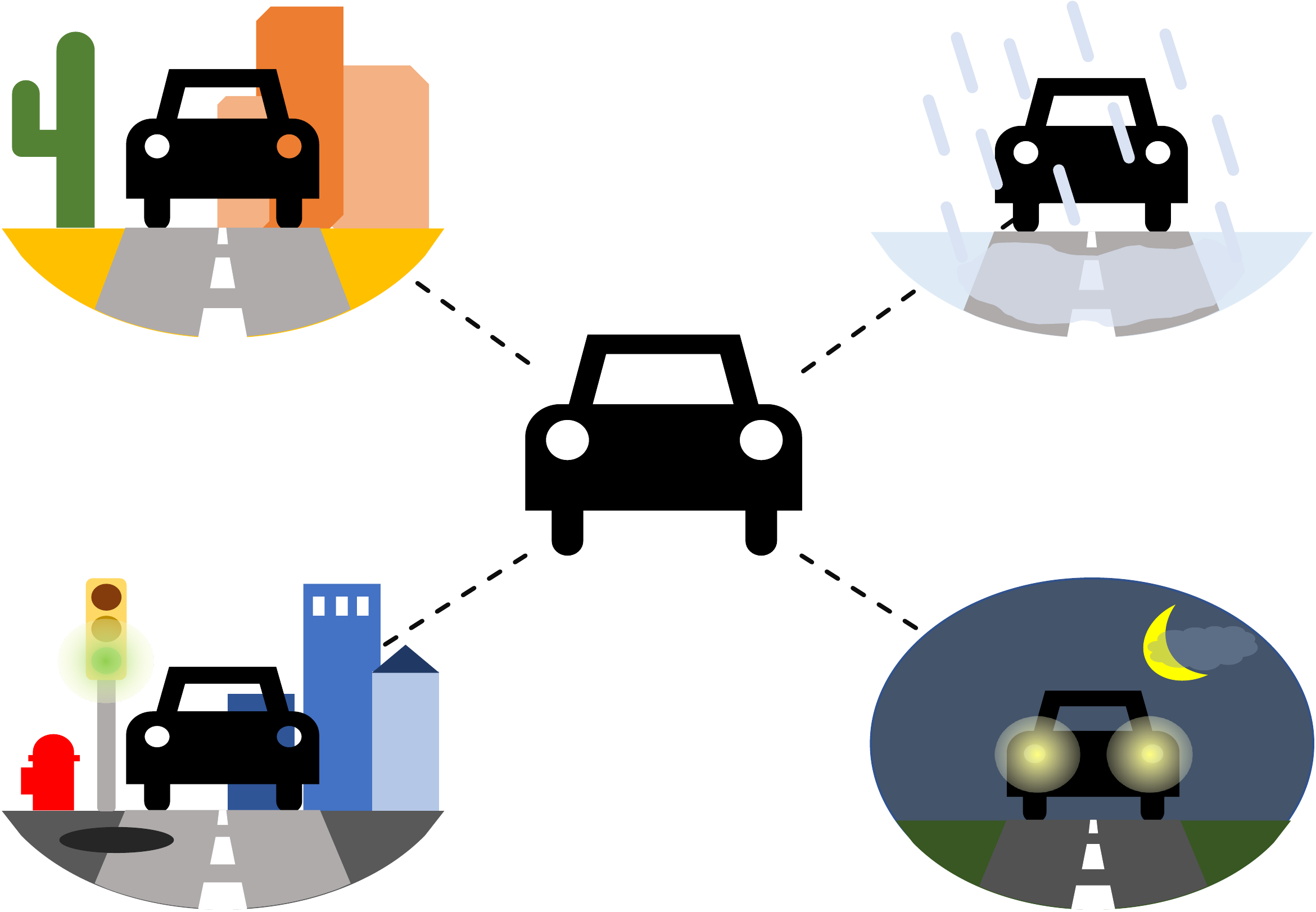}
        \caption{Self-driving}
        \label{fig:auto-driving}
    \end{subfigure}
    \hfill
    \begin{subfigure}[b]{0.43\textwidth}
    \centering
        \includegraphics[width=\textwidth]{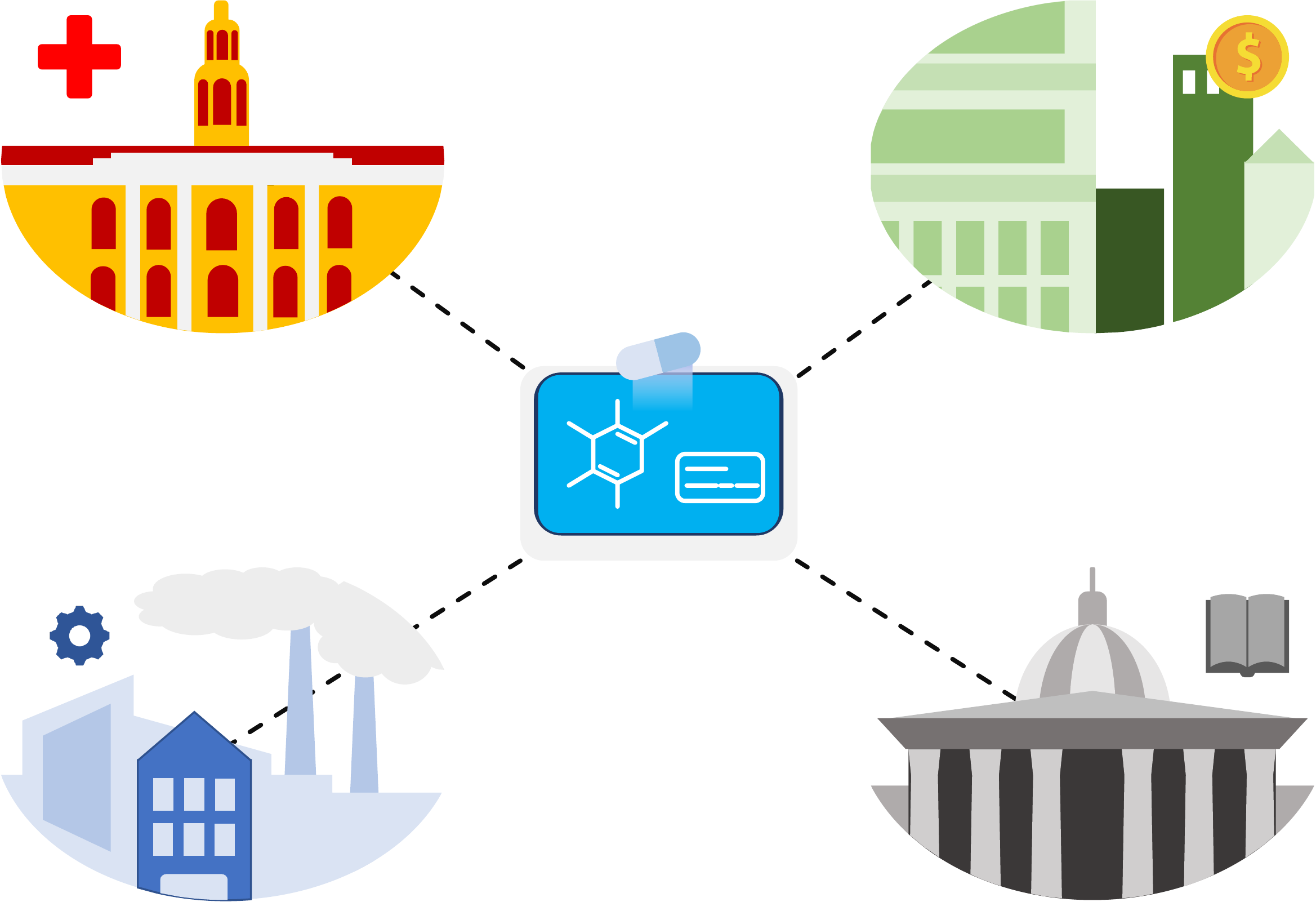}
        \caption{Drug discovery}
        \label{fig:drug-discovery}
    \end{subfigure}
    \hspace{20pt}
    \vskip\baselineskip
    \hspace{20pt}
    \begin{subfigure}[b]{0.43\textwidth}
    \centering
        \includegraphics[width=\textwidth]{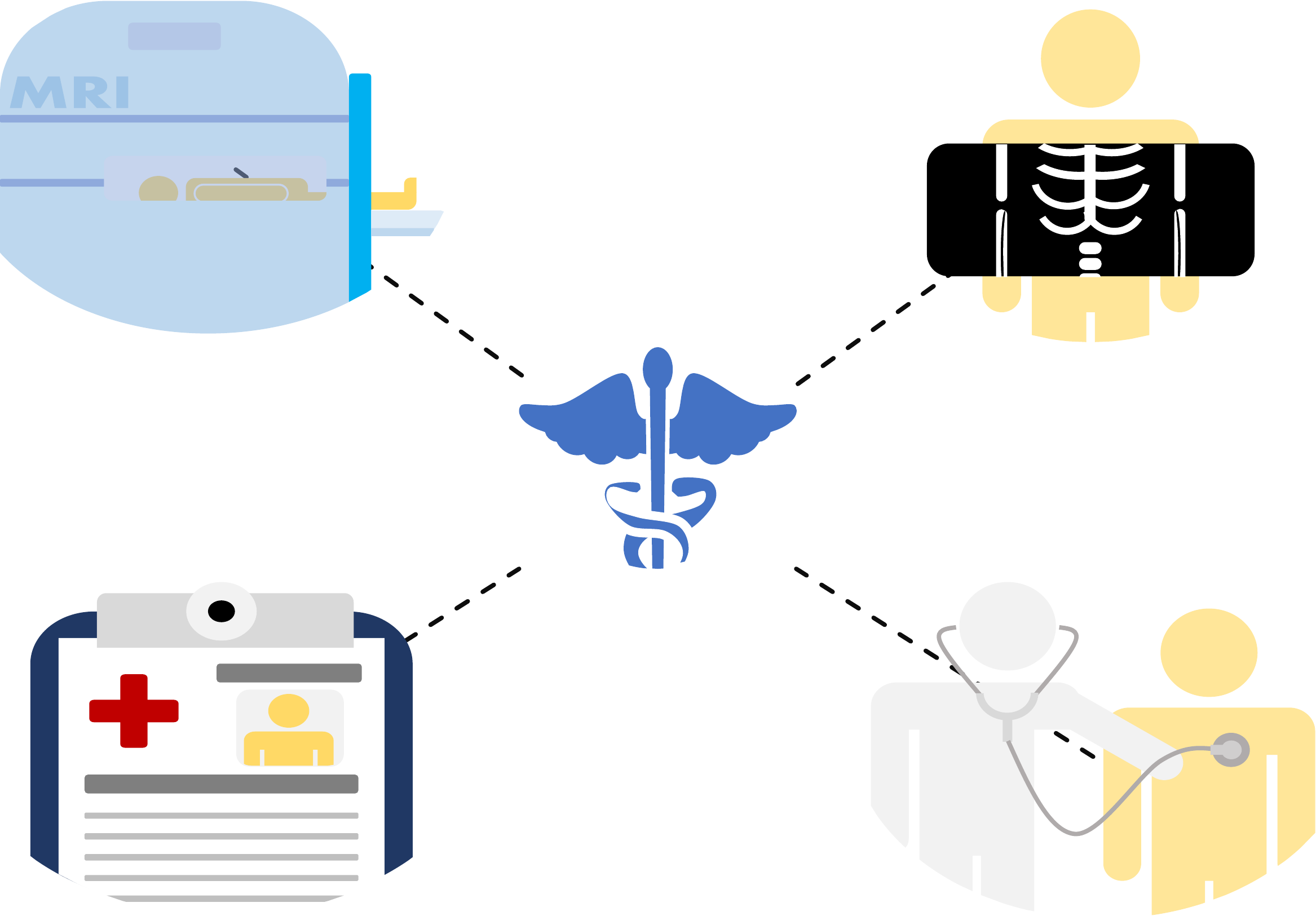}
        \caption{Clinical diagnosis}
        \label{fig:clinical-diagnosis}
    \end{subfigure}
    \hfill
    \begin{subfigure}[b]{0.43\textwidth}
    \centering
        \includegraphics[width=\textwidth]{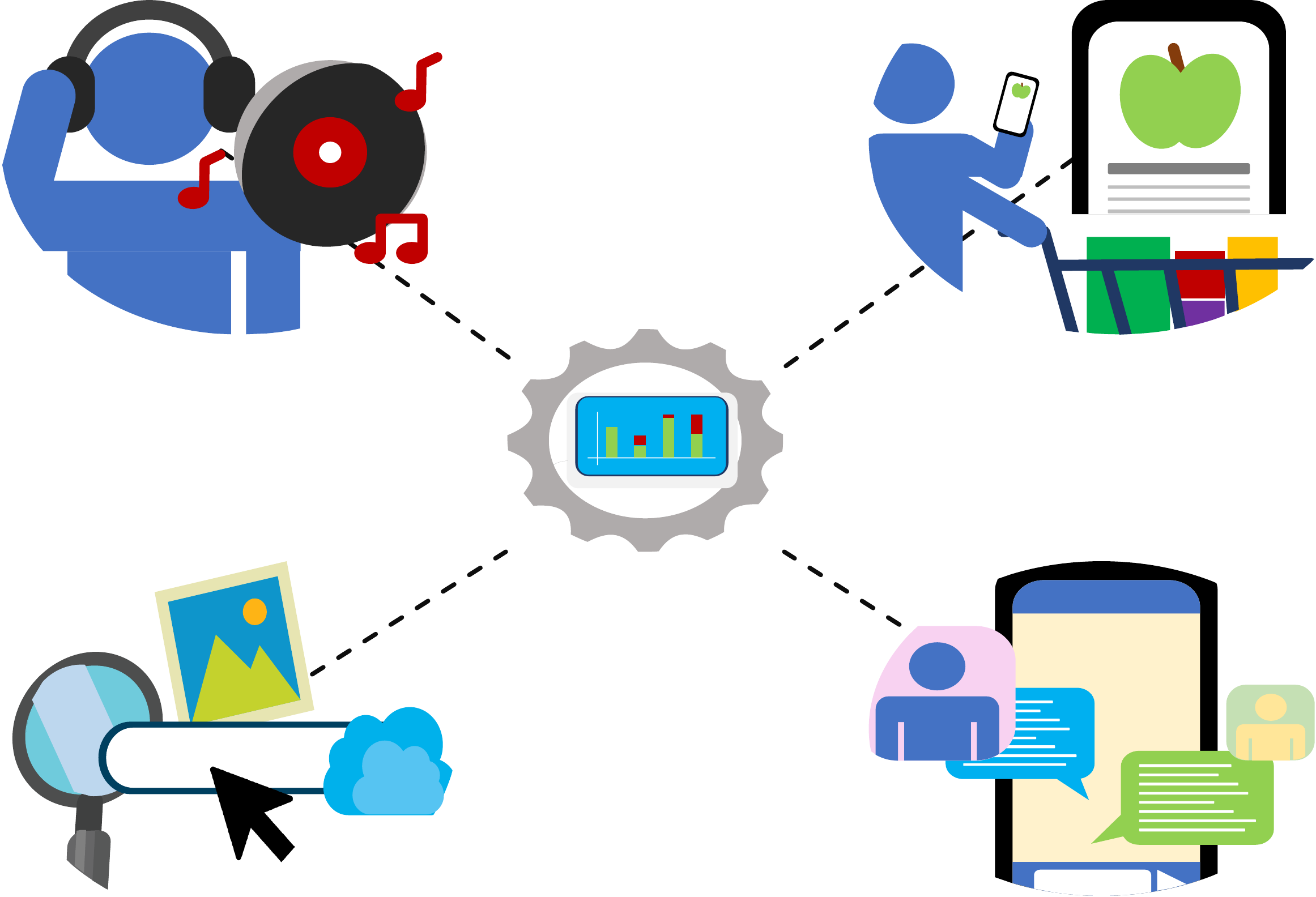}
        \caption{Recommendation}
        \label{fig:recommendation-system}
    \end{subfigure}
    \hspace{20pt}
    \caption{Real-life use-cases of FL. In (a), a better self-driving system can be trained by combining the varied experiences of individual vehicles across the full diversity of road scenes. In (b), FL enables pharmaceutical companies to collaborate in an otherwise highly competitive space, thereby accelerating drug discovery. In (c), more accurate clinical diagnoses can be made by considering multiple modalities of information. In (d), FL can make for higher quality recommendations through on-device personalization.}
    \label{fig:use_case}
\end{figure*}

FL has already been employed in many different applications; here we discuss a few representative scenarios, as illustrated in Figure~\ref{fig:use_case}. 
These scenarios also motivate OpenFed's support for third-party libraries commonly used in industry. 
Specifically, OpenFed-CV and OpenFed-RL address the perception and decision-making components respectively for self-driving systems. 
OpenFed-Medical supports common medical use cases. 
Finally, OpenFed-NLP can be used for the text-based recommendation systems described.

\paragraph{Self-driving}
Self-driving systems can typically be broken down into two major components: perception, and decision-making~\cite{aledhari2020federated}. 
The perception system is generally further divided into subsystems responsible for self-localization, static obstacle mapping, moving obstacle detection and tracking, road mapping, and traffic signal detection, and recognition, among others. 
The decision-making system likewise comprises tasks such as route planning, path planning, behavior selection, motion planning, and control. 
These tasks are highly complex and each requires not only large volumes of data but also data of a sufficiently wide diversity to account for the natural diversity of street scenes. 

A promising source of data is the fleet of vehicles already sold - in the course of their everyday operation, these vehicles would collectively accumulate data approaching the scale and variety needed. 
Due to privacy concerns and the high costs of transmitting rich multimedia data, the traditional model of data aggregation and centralized training is unfeasible.
FL addresses these problems directly - for example, a previous case study demonstrated that using FL for the steering wheel prediction problem reduces bandwidth by 60\% and training time by 70\% with no loss to the model performance~\cite{zhang2021real}.

\paragraph{Drug discovery}
According to Pharmaceutical Research and Manufacturers of America, it takes on average 10 years and \$2.6 billion for a new medicine to reach the market~\cite{phrma}. 
One part of this process that machine learning can accelerate is drug discovery. 
For example, quantitative structure-activity relationship (QSAR) is a machine learning method for predicting the relationship between chemical structures and resultant biological activities.
One challenge in improving the performance of QSAR models is data availability. Data from different institutions cannot be freely shared and aggregated due to commercial and legal reasons. FL has been demonstrated to work well for QSAR~\cite{xiong2020facing,chen2020fl}, and can overcome the data availability problem.

Furthermore, FL for drug discovery has moved beyond the theoretical. 
The MELLODDY project is a federated learning platform to accelerate drug discovery for 10 major pharmaceutical companies, allowing them ``for the first time to collaborate in their core competitive space''~\cite{mellody}.

\paragraph{Clinical diagnosis}
Clinical diagnoses are often made with context provided by a diverse range of sources, from patient medical history to different types of imagery. 
This use of multi-modal information is required to give a holistic assessment of the patient's condition. 
Vertical FL, also known as Heterogeneous FL, allows different types of information across different data sources to be combined securely and with privacy protection. 
This is critical for such a system to be deployed in practice, as medical data and records are highly sensitive. 

Previous work has demonstrated the need for more diverse data in this clinical diagnosis setting~\cite{ziller2021medical}. 
Medical data can be highly heterogeneous across institutions - it was found that models trained using data from single institutions perform substantially worse on test examples from other institutions. 
By using FL, models can benefit from a greater diversity of cases and thereby perform better even on validation data from their own institution~\cite{2020Federated}.

\paragraph{Recommendation} 
FL has been successfully applied to recommendation systems such as browser history suggestion~\cite{hartmann2019federated}, keyboard query suggestion~\cite{yang2018applied}, and mobile keyboard prediction~\cite{leroy2019federated}. 
The mobile keyboard prediction study provides a rare real-world large-scale comparison between an FL system and the best-performing centralized alternative. 
The centralized model benefited from the relative maturity of its traditional machine learning setup - including best practices for the algorithm design, parameter selection, and training methodology. 
However, it was limited to users who had opted in and allowed their data to be recorded. 
In this setup, it was found that the FL model did not simply match the centralized model performance, but in fact, exceeded it in terms of both the top-1 and top-3 prediction metrics by 1\%.

\section{Discussion}

We present OpenFed, a comprehensive and versatile open-source framework for FL with a variety of benchmarks across diverse federated tasks. 
In this section, we further discuss the strengths and weaknesses of OpenFed as well as future work directions from the perspective of facilitating academic and industrial research and development.

\paragraph{Comparison with prior FL frameworks}
As shown in Table \ref{tab:platform-summary}, although there exist other FL frameworks, most of them focus on specific uses and are not meant for general purposes. 
A gap exists for a framework to accelerate underlying FL algorithm research. 
OpenFed bridges this gap. The main advantages of OpenFed compared to existing frameworks are summarized as follows:
\begin{itemize}
    \item \textbf{Inherited design structure.} In general, most existing frameworks are based on either TensorFlow or PyTorch. However, they typically do not follow the programming structure of their underlying frameworks. Instead, they propose their own proprietary designs. Uniquely, our implementation completely follows the design philosophy of our base, PyTorch. The relatively large DL community should thus find OpenFed to be immediately familiar. Researchers using OpenFed can thus concentrate on exploring and implementing novel algorithms, while developers will find it easy to apply these algorithms to their tasks.
    \item \textbf{Modular algorithm implementations. }Most frameworks support only the simplest FL optimization algorithm, FedAvg~\cite{kairouz2019advances,jiang2020optimal}. Although some frameworks also include other more complex algorithms, the implementations tend to be more functional. In comparison, the OpenFed design clearly demarcates where each step of the FL workflow begins and ends. We summarize and standardize the workflow by describing it with four independent steps, i.e. parameter aggregation, gradient accumulation, parameter penalization, and state synchronization. Different FL algorithms can thus be clearly implemented by making specific adjustments within only the required steps. This philosophy not only facilitates researchers to realize standardized and maintainable code but also makes explicit where the differences are between different algorithms.
    \item \textbf{Flexible topology support.} Most frameworks support only centralized topologies and provide limited support for some of the more complex ones. Research on different topologies is an important topic in FL. Our concept of the federated group allows OpenFed to flexibly implement any topology.
\end{itemize}

\paragraph{Future work} As FL continues to gain traction, we anticipate that new features will become necessary for OpenFed to continue to meet the requirements of academia and industry.
\begin{itemize}
    \item \textbf{Topology benchmarks.} At present, all benchmarks default to the centralized topology for algorithm evaluation, and no relevant large-scale experiments have been conducted on the impact of different topologies on FL performance. We will continue to promote this work as OpenFed matures.

    \item \textbf{Industry deployment.} OpenFed runs in a Python runtime environment, which makes it challenging for deployment on resource-constrained devices. Therefore, more efficient runtime environments such as C/C++ must be supported. Furthermore, we will continuously increase the number of third-party libraries that OpenFed supports~\cite{bonawitz2019towards}.

    \item \textbf{Cryptography.} Although there are many mature symmetric/asymmetric encryption algorithms in the communication field, these algorithms are often too complex to encrypt large amounts of data and consume too much computation. An open problem is therefore to design an effective encryption algorithm under the condition of limited computing resources according to the characteristics of FL.

    \item \textbf{Poisoning and adversarial attacks.} The data used for training in FL often involves personal and private information. There has been researched work demonstrating vulnerabilities in FL, for example, it has been shown that the content of training data can be inferred (to some degree) through the exposed gradients~\cite{yazdinejad2020p4,kim2019blockchained}. OpenFed requires more defense mechanisms and algorithms to counteract these attacks.
\end{itemize}

\paragraph{Limitations.}
With the growing importance of privacy protection and data security, FL has become increasingly adopted across different fields. 
However, there is still generally a performance gap between FL and traditional distributed learning. 
As such, in scenarios where the data is not particularly sensitive, researchers and developers still prefer traditional distributed learning. 
We hope that OpenFed can address the pain points of FL research and stimulate the enthusiasm of researchers new and experienced, ultimately narrowing the performance gap.

\section{Conclusion}
We present a free, open-source software framework for FL and end-to-end encrypted inference, which we showcased in several relevant real-life case studies. 
OpenFed is comprehensive and versatile and can be used both for research and development as well as for commercial and industrial applications. 
Ongoing work will enable the large-scale deployment of OpenFed, the validation of our findings on diverse cross-institutional datasets, and further the widespread utilization of our framework.


{\small
\bibliographystyle{ieee_fullname}
\bibliography{egbib}
}

\end{document}